\documentclass[pdftex,twocolumn,epjc3]{svjour3}          

\RequirePackage[T1]{fontenc}

\smartqed  

\RequirePackage{graphicx}
\RequirePackage{mathptmx}      
\RequirePackage{flushend}
\RequirePackage[numbers,sort&compress]{natbib}
\RequirePackage[colorlinks,citecolor=blue,urlcolor=blue,linkcolor=blue]{hyperref}
\usepackage{amsmath}

\journalname{Eur. Phys. J. C}

\begin{document}

\title{Distinguishing $\Lambda$CDM from evolving dark energy with the future gravitational-wave space-borne detector DECIGO}

\author{Yuting Liu\thanksref{addr1,addr2,addr3}
       \and
        Shuo Cao\thanksref{e1,addr1,addr2}
        \and
        Xiaogang Zheng\thanksref{addr4}
        \and
        Marek Biesiada\thanksref{addr5}
        \and
        Jianyong Jiang\thanksref{e2,addr2}
        \and
        Tonghua Liu\thanksref{addr6} 
}

\thankstext{e1}{e-mail:caoshuo@bnu.edu.cn}
\thankstext{e2}{e-mail:jianyong@bnu.edu.cn}

\institute{Institute for Frontiers in Astronomy and Astrophysics, Beijing Normal University, Beijing 102206, China\label{addr1}
          \and
         School of Physics and Astronomy, Beijing Normal University, Beijing 100875, China\label{addr2}
          \and
          Department of Physics, University of Tokyo, Tokyo 113-0033, Japan\label{addr3}
          \and
          School of Electrical and Electronic Engineering,Wuhan Polytechnic University, Wuhan 430023, China\label{addr4}
          \and
          National Centre for Nuclear Research, Pasteura 7, 02-093 Warsaw, Poland\label{addr5}
          \and
          School of Physics and Optoelectronic, Yangtze University, Jingzhou 434023, China\label{addr6}
}

\date{Received: date / Accepted: date}

\maketitle

\begin{abstract}
The $Omh^2(z_i,z_j)$ two point diagnostics was proposed as a litmus test of $\Lambda$CDM model and measurements of cosmic expansion rate $H(z)$ have been extensively used to perform this test. The results obtained so far suggested a tension between observations and predictions of the $\Lambda$CDM model. However, the dataset of $H(z)$ direct measurements from cosmic chronometers and BAO was quite limited. This motivated us to study the performance of this test on a larger sample obtained in an alternative way. In this Letter, we propose that gravitational wave (GW) standard sirens could provide large samples of $H(z)$ measurements in the redshift range of $0<z<5$, based on the measurements of dipole anisotropy of luminosity distance arising from the matter inhomogeneities of large-scale structure and the local motion of observer. We discuss the effectiveness of our method in the context of the future generation space-borne DECi-herz Interferometer Gravitaional-wave Observatory (DECIGO), based on a comprehensive $H(z)$ simulated data set from binary neutron star merger systems. Our result indicate that in the GW domain, the $Omh^2(z_i,z_j)$ two point diagnostics could effectively distinguish whether $\Lambda$CDM is the best description of our Universe. We also discuss the potential of our methodology in determining possible evidence for dark energy evolution, focusing on its performance on the constant and redshift-dependent dark energy equation of state.
\end{abstract}

\section{Introduction}

One of the most important challenges in modern cosmology is to understand the fundamental nature of the 
concordance cosmological model, which include three pillars: inflation (flat universe), dark matter (neutral, collisionless particles) and dark energy (cosmological constant) \citep{Guth1981,Turner1984,Lyth1999,Peebles2003,Bertone2005,Turner2018}. Since the 1980s, the concordance model has withstood many rigorous tests based on different astronomical observations such as type Ia supernovae (SNe Ia) \citep{Riess1998,Perlmutter1999}, cosmic microwave background (CMB), baryon acoustic oscillations (BAO) \citep{Spergelc2003,Eisenstein2005}, and strong gravitational lensing \citep{CaoAPJ2012,CaoJCAP2012,CaoAPJ2015}. However, this harmony has recently been disrupted by the increasing precision of available measurements. Currently, the most hotly debated issue is the ``$H_0$ tension'', where the value of Hubble constant measured with CMB ($H_0=67\pm0.5$ $km\; s^{-1}Mpc^{-1}$) based on the $\Lambda$CDM assumption is (4 -- 6)$\sigma$ different from the supernovae direct local measurement ($H_0=73\pm1.7$ $km\;s^{-1}Mpc^{-1}$) \citep{Riess2019,Verde2019}. At the same time, another tension emerges regarding cosmic curvature $\Omega_k$. The CMB anisotropies analysed alone tend to support a closed universe, which is 3$\sigma$ different from the results obtained jointly with BAO data \citep{DiValentino2020,Handley2021}. In addition, there is a discrepancy regarding the matter density parameter $\Omega_m$, which is noticeably smaller in recent cosmic shear surveys than the value obtained from \textit{Planck} data in the framework of $\Lambda$CDM \citep{Asgari2020}. It is also worth noting that the 4$\sigma$ difference exists between the $\Lambda$CDM and the high-redshift Hubble diagram of the type Ia supernovae (SNe Ia), quasars (QSOs), and gamma-ray bursts (GRBS) \citep{Risaliti2019,Lusso2019}. Although systematic uncertainties might play an important role in these tensions, they may also indicate that $\Lambda$CDM model is not valid when confronted with more accurate data. As a result, it is natural and necessary to re-examine the performance of the $\Lambda$CDM, because its failure could lead to new discoveries that may give us even greater surprises. This has prompted and motivated many researchers to test and discuss the foundations of the standard cosmological model \citep{Yang2020,DiValentino2021,Vagnozzi2021,Koksbang2021}.

Two point diagnostics $Omh^2(z_1,z_2)$ developed by \citet{Sahni2014} is an attractive technique used to test the $\Lambda$ Cold Dark Matter model, using the measurements of Hubble parameter $H(z)$ derived from observations. It is more convenient and alleviates some issues with the original one point diagnostic $Om(z)$, which has also been widely implemented with the $H(z)$ measurements obtained from cosmic chronometers and baryon acoustic oscillations \citep{Ade2014,Ding2015,Zheng2016,Qi2018,Cao2018}.
However, the number of measured $H(z)$ is still too low to guarantee a sufficient statistical power. Moreover, the redshift coverage is limited to $z\sim 2$. Robust reconstruction techniques such as Gaussian Process are a common approach to alleviate these drawbacks. It should be noted that the reconstruction may produce different systematic errors, which will naturally affect the reliability of the results. Fortunately, the detection of gravitational wave (GW) signals has opened a new window on the Universe and provides more possibilities \citep{Schutz1986,LIGO2017}. Methods of directly measuring the expansion rate $H(z)$ using gravitational waves have also been proposed and discussed \citep{Sasaki1987,Seto2001,Bonvin2006a,Bonvin2006b,Nishizawa2011}. This is exciting because it means that we have more chances to get a larger sample of $H(z)$ across a higher redshift range to test the consistency of $\Lambda$CDM with two point diagnostics $Omh^2(z_1,z_2)$ technique.

In this paper, we will focus on this method of testing the consistency of $\Lambda$CDM \citep{Sahni2014}, using the Hubble parameter $H(z)$ from the simulated sample of data attainable to GW detector DECIGO up to $z\sim$ 5. Aiming at the verification of $\Lambda$CDM in both electromagnetic (EM) and gravitational wave domain, we also discuss the currently largest 32 model-independent measurements of the Hubble parameter $H(z)$. Moreover, we investigate the performance of $Omh^2(z_1,z_2)$ two-point diagnostics under $\omega$CDM and CPL models. In Sect.~\ref{sec:data}, we briefly describe the $Omh^2(z_1,z_2)$ two point diagnostics method and $H(z)$ samples in GW domain and EM domain. In Sec.~\ref{sec:result}, the results and discussions are presented. The conclusions are summarized in Sec.~\ref{sec:con}.

\section{Methodology and Data}\label{sec:data}

Since its inception, the $Omh^2(z_1,z_2)$ diagnostics has developed into an interesting and popular method extending the repository of tools and techniques of model testing and comparison.
By definition, the $Omh^2(z_i,z_j)$ two-point diagnostics can be written as
\begin{equation} \label{eq:Omh2}
	Omh^2(z_i,z_j) = \frac{h^2(z_i)- h^2(z_j)}{(1+z_i)^3 - (1+z_j)^3},
\end{equation}
where $h(z) \equiv H(z)/100\;km\,s^{-1}\,Mpc^{-1}$ \citep{Sahni2014}.
It means that once we get the measurements of the Hubble parameter $H(z)$ at two or more redshifts directly from the observations, we can obtain the $Omh^2(z_i,z_j)$ values and further compare with its theoretical values derived from concrete cosmological models \citep{LiuPLB2023}.

In the simplest case of the $\Lambda$CDM model, theoretical expression of Eq.~(\ref{eq:Omh2}) can be written as follows
\begin{equation}
	Omh^2(z_i,z_j)_{\Lambda CDM}=\Omega_{m,0}h^2,
\end{equation}
where $h \equiv H_0 / 100\;km\,s^{-1}\,Mpc^{-1}$. 
A significant advantage of the $Omh^2(z_i,z_j)$ two-point diagnostics for $\Lambda$CDM is that once some other probe tells us the value of $\Omega_{m,0}h^2$, we are able to know the value of $Omh^2(z_i,z_j)$. As a result, any departure of $Omh^2(z_i,z_j)$ from the above value may indicate that dark energy is not $\Lambda$. In other words, we can use this deviation to test $\Lambda$CDM model and distinguish it from other dark energy models. In the case of other cosmological models, the two-point diagnostics is no longer a single number, but its theoretical counterpart can be formulated. The simplest type of a model different from $\Lambda$CDM is the dynamical dark energy $\omega$CDM model with a constant equation of state $p = \omega \rho$. The two-point diagnostics in this model can be expressed as
\begin{align} \label{eq:Omh2_wCDM}
	Omh^2(z_i,z_j)_{(wCDM)} = & \Omega_{m,0}h^2+(1-\Omega_{m,0})h^2 \nonumber \\
	& \left[\frac{(1+z_i)^{3(1+w)} - (1+z_j)^{3(1+w)}}{(1+z_i)^3-(1+z_j)^3}\right].
\end{align}
In addition, we also consider the Chevallier-Polarski-Linder (CPL) evolving dark energy model \citep{Chevalier2001,Linder2003} with the equation of state $p = w(z) \rho$, where $w(z) = w_0 + w_a \frac{z}{1+z}$. In this framework, the theoretical expression of Eq.~(\ref{eq:Omh2}) can be expressed as
\begin{align} \label{eq:Omh2_CPL}
	Omh^2(z_i,z_j)_{(CPL)} = & \Omega_{m,0}h^2 + (1-\Omega_{m,0})h^2 \nonumber \\
	& \biggl[( (1+z_i)^{3(1+w_0+w_a)}e^{\frac{- 3 w_a z_i}{1+z_i}} \nonumber \\
	& - (1+z_j)^{3(1+w_0+w_a)} e^{\frac{-3w_a z_j}{1+z_j}})
	\nonumber \\
	& /((1+z_i)^3 - (1+z_j)^3)\biggr].
\end{align}
Combined with Eq.~(\ref{eq:Omh2}), it becomes obvious and natural to test the $\Lambda$CDM model and differentiate the cosmological model, based on the measurements of the Hubble parameter $H(z)$.

\begin{figure}
	\begin{center}
		\includegraphics[width=0.95\linewidth]{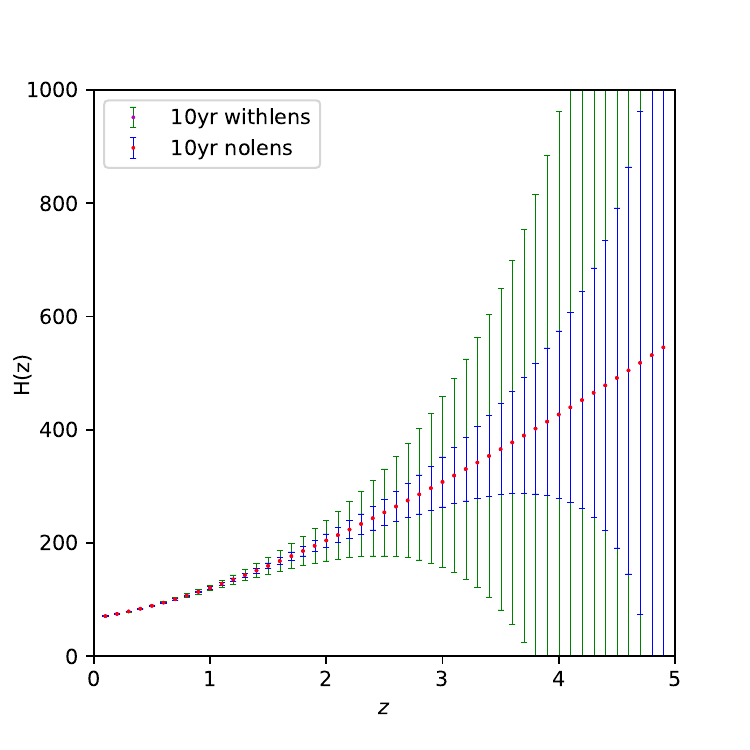}
	\end{center}
	\caption{Simulated $H(z)$ data from DECIGO. Blue bars with smaller uncertainty represent cases without lensing effect while the green ones take the lensing effect into account. }
	\label{Fig.sHz}
\end{figure}

\subsection{Hubble parameter $H(z)$ from GW standard sirens detectable by DECIGO}\label{subsec:Hz_DECIGO}

DECi-hertz Interferometer Gravitational-wave Observatory (DECIGO) as a future space-borne gravitational-wave antenna will be sensitive to lower frequency range $f= 0.1-10\;Hz$ and will have higher detection sensitivity than ground based interferometric detectors \citep{Seto2001,Kawamura2011,Kawamura2019}. It would be a new opportunity to detect signals of gravitational wave sources (neutron star binaries and black hole binaries) in the inspiral phase much earlier. These binary systems would enter the sensitivity band of ground-based detectors up to a few years later. In particular, DECIGO will be able to detect the GW from neutron star binaries even at a redshift of z$\sim$5 for five years of its mission. It is well established that coalescing binaries (in the inspiral phase) are the standard sirens and can provide direct measurements of the luminosity distance. This opportunity has been widely discussed \citep{Schutz1986,Qi2019,Geng2020}. As a matter of fact, it also opens up more possibilities: such as probing the spatial geometry of the Universe \citep{Zheng2021,CaoAPJ2022,ZhangAPJ2022}, dark photon bursts from compact binary systems \citep{HouPRD2022}, direct tests of the FLRW metric \citep{CaoSR2019}, the distribution of dark matter in the Universe \citep{CaoMN2021,CaoAA2022}, and measuring Hubble parameters based on the dipole anisotropy of luminosity distance from GW observations \citep{Qi2021}.

Considering that matter in the large-scale structure of the Universe is not strictly uniformly distributed and the coordinate system we use is not at absolute rest, the real luminosity distance (neglecting multipoles higher than the dipole) needs to be redefined as \citep{Bonvin2006a,Bonvin2006b}
\begin{equation}
	d_L(z,\mathbf{n})=d_L^{(0)}(z)+d_L^{(1)}(z)(\mathbf{n}\cdot\mathbf{e}),
\end{equation}
where $z$ is the redshift and $\mathbf{n}$ is the direction to the source. The $d_L^{(0)}(z)$ means the luminosity distance to a source in an unperturbed Friedman Universe, in other words, it is the luminosity distance in the traditional sense. Here $\mathbf{e}=\mathbf{v_0}/|\mathbf{v_0}|$ is a unit vector in the direction of the dipole and $d_L^{(1)}(z)$ as its amplitude can be represented in the following form:
\begin{equation}
	d_L^{(1)}(z)=\frac{|\mathbf{v_0}|(1+z)}{H(z)}.
\end{equation}
In principle, $\mathbf{v_0}$ can be directly measured from the CMB dipole \citep{Nishizawa2011} and its magnitude is $v_0 = 369.1\pm{0.9}\;km\,s^{-1}$ based on  the COBE satellite measurements. Consequently, we can get the Hubble parameters $H(z)$ directly from the dipole of luminosity distance with the uncertainty assessed in the following way \citep{Bonvin2006a,Bonvin2006b,Nishizawa2011} 
\begin{equation}
	\frac{\Delta H(z)}{H(z)}=\frac{\Delta d_L^{(1)}(z)}{d_L^{(1)}(z)}=\sqrt{3}\left[\frac{d_L^{(1)}(z)}{d_L^{(0)}(z)}\right]^{-1}\left[\frac{\Delta d_L^{(0)}(z)}{d_L^{(0)}(z)}\right].
\end{equation}
The relative uncertainty $\Delta d_L^{(0)}(z)/d_L^{(0)}(z)$ includes instrumental uncertainty $(\sigma_{inst})$ caused by the GW detector DECIGO and other systematic uncertainties caused by the gravitational lensing $(\sigma_{lens})$ and the peculiar velocity $(\sigma_{pv})$. Namely,
\begin{equation}\label{eq7}
	\left[\frac{\Delta d_L^{(0)}(z)}{d_L^{(0)}(z)}\right]^2=\sigma_{inst}^2+\sigma_{lens}^2+\sigma_{pv}^2.
\end{equation}
One can estimate instrumental uncertainty $(\sigma_{inst})$ by Fisher matrix method, for more details one can refer to \citet{Nishizawa2011}. Concerning the $\sigma_{lens}$ uncertainty induced by the matter inhomogeneities of large-scale structure along the line of sight, it can be estimated as \citep{Hirata2010}
\begin{equation}
	\sigma_{lens}(z)=0.066\left[\frac{1-(1+z)^{-0.25}}{0.25}\right]^{1.8}.
\end{equation}
In addition, the systematic uncertainty caused by Doppler effect induced by the peculiar velocity of the source along the line of sight is given by \citep{Gordon2007}
\begin{equation}
	\sigma_{pv}(z)=\left|1-\frac{(1+z)^{2}}{H(z)d_L^{(0)}(z)}\right|\sigma_{v,gal}.
\end{equation}
Usually, the 1-dimensional velocity dispersion of the galaxy $\sigma_{v,gal} = 300$km/s can be assumed \citep{Silberman2001}.

One thing worth noting is that we focus on the dipole anisotropy of luminosity distance from the gravitational waves produced by the binary neutron stars (BNSs). Therefore, for the BNS merge rate we employ the fitting form of the redshift distribution of GW sources obtained from the cosmic star formation \citep{Schneider2001,Cutler2009,Sathyaprakash2010}. Based on the detection rate of $10^4$ neutron star binaries \citep{Kawamura2021}, we simulated the total number of $10^5$ BNS as a representative yield of DECIGO 10-year mission. Considering all of the total uncertainties in Eq.~(\ref{eq7}), we display the simulated Hubble parameter $H(z)$ obtained from DECIGO in Fig.~\ref{Fig.sHz} (with the redshift bin of $\Delta z=0.1$). The fiducial cosmology adopted in the simulation is the $\Lambda$CDM model with the Planck 2018 results \citep{Asgari2020}. In addition, considering that the systematic uncertainty caused by line-of-sight lensing may be eliminated by some feasible techniques \citep{Hirata2010,Shapiro2010}, we also consider the cases without the lensing effect for comparison. 
In Fig.~\ref{Fig.sHz}, we plot the errors of $H(z)$ taking account of the case with and without lensing errors, during 1-year and 10-year observations. Our results show that the uncertainty of $H(z)$ over 10-year observations is significantly smaller than that for 1-year observation, proportional to $1/\sqrt{N(z)}$. Here $N(z)$ is the number of independent BNS systems in the vicinity of the redshift $z$. 


\section{Results and Discussion}\label{sec:result}

Having obtained Hubble parameters $H(z)$ from the simulated standard sirens and taking the measurements of cosmic chronometers, we can calculate the $Omh^2(z_i,z_j)$ values of every combination of $H(z)$ pairs. When we take a look at the two-point diagnostics with their uncertainties as a function of redshift difference $\Delta{z}=|z_i-z_j|$. We could see some interesting features regarding the uncertainties. Namely, as the observation time increases, the uncertainty decreases substantially, regardless of whether we consider the lensing effect or not. In other words, accuracy will improve dramatically with 10 year observations. Compared with the case without lensing error, the results accuracy of $Omh^2(z_i,z_j)$ are degraded. Two point diagnostics as tests of the $\Lambda$CDM model provide numerical values for each combination of data pairs, which should in principle reproduce a single number. Hence, we need to analyze these individual values from a statistical point of view.

\begin{figure*}
	\begin{center}
		\includegraphics[width=0.65\linewidth]{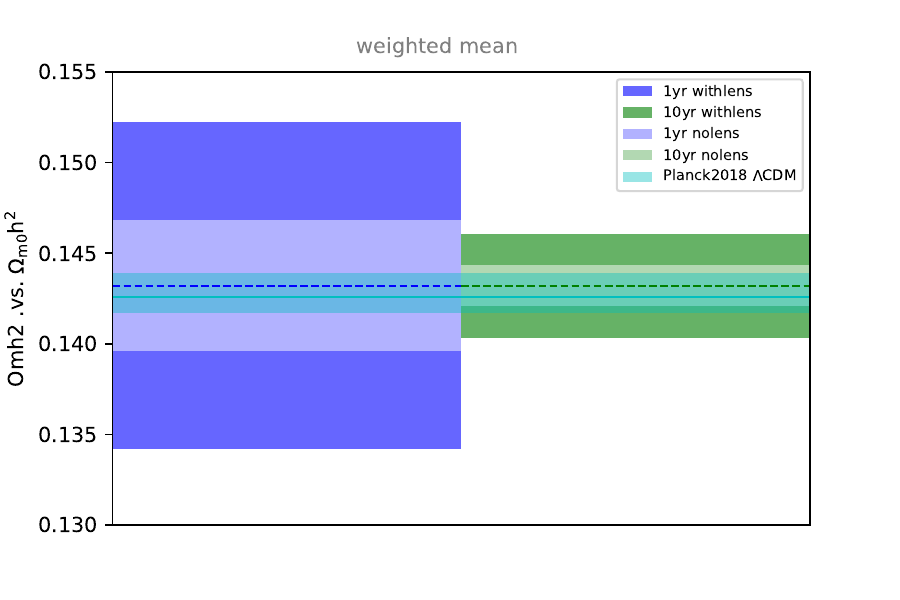}
		\caption{The weighted mean of the $Omh^2(z_i,z_j)$ two point diagnostics displayed by dashed lines and surrounded by color bands denoting 1$\sigma$ confidence level. The four different color lines and bands show results with and without lensing effect in 1-year and 10-year observations, respectively. The cyan line and band are the Planck2018 result.}
		\label{Fig3}
	\end{center}
\end{figure*}

\begin{table*}
	\centering
	\caption{The Weighted Mean (W.M.) of $Omh^2(z_i,z_j)$ Two-point Diagnostics and  Residuals calculated for $\Lambda$CDM, $\omega$CDM and CPL using the simulated $H(z)$ data.
	}
	\begin{tabular}{c c c c c} \hline\hline
		&   $Omh^2(z_i,z_j)_{(w.m.)}$   & $R_{(w.m.)(\Lambda CDM)}$  & $R_{(w.m.)(\omega CDM)}$ & $R_{(w.m.)(CPL)}$  \\
		\hline
		$H(z)_{1yr+withlens} $  & $0.143\pm{0.009}$  &  $4.46\times10^{-6}\pm0.0090$  &  $0.0044\pm0.0091$  &  $0.0058\pm0.0092$ \\
		$H(z)_{10yr+withlens} $  & $0.143\pm{0.003}$ &  $5.00\times10^{-6}\pm0.0029$  &  $0.0043\pm0.0030$  &  $0.0054\pm0.0033$ \\
		$H(z)_{1yr+withoutlens}$  & $0.143\pm0.004$ & $1.57\times10^{-5}\pm0.0036$   &  $0.0036\pm0.0037$  &  $0.0049\pm0.0038$ \\
		$H(z)_{10yr+withoutlens}$ & $0.143\pm0.001$ &  $1.80\times10^{-5}\pm0.0012$   &  $0.0033\pm0.0013$  &  $0.0041\pm0.0015$ \\
		\hline\hline
	\end{tabular}
	\label{Table1}
\end{table*}

\begin{figure*}
	\begin{center}
		\includegraphics[width=8cm,height=6.5cm]{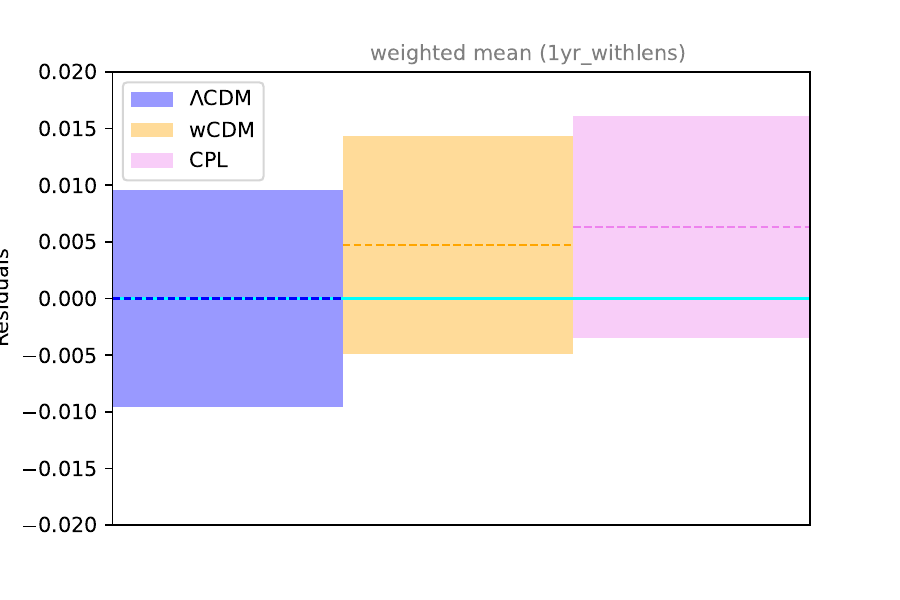}
		\includegraphics[width=8cm,height=6.5cm]{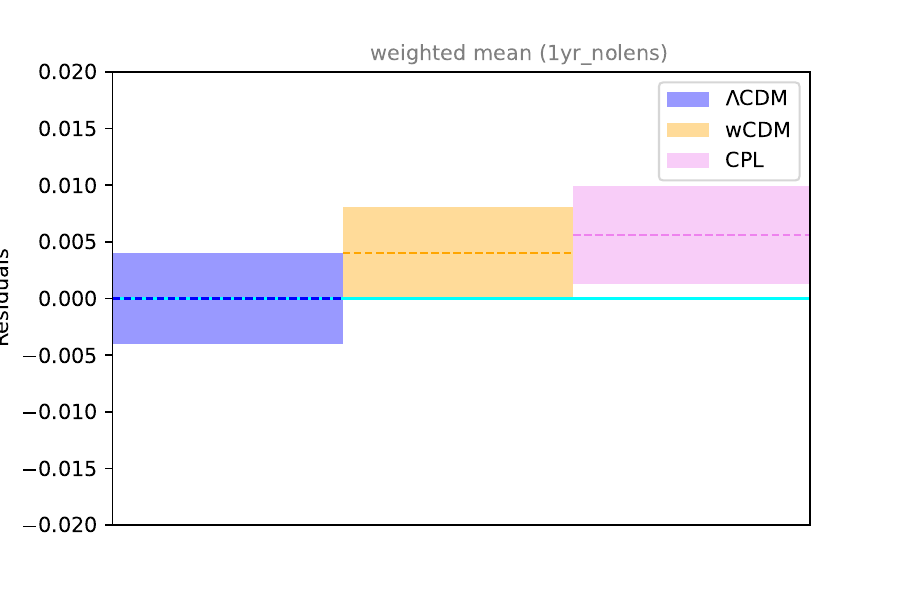}
		\includegraphics[width=8cm,height=6.5cm]{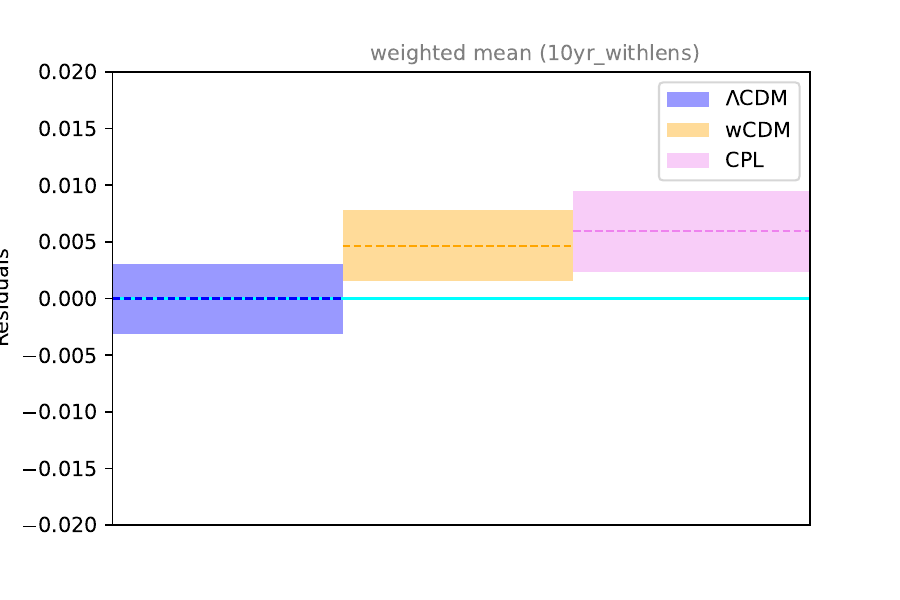}
		\includegraphics[width=8cm,height=6.5cm]{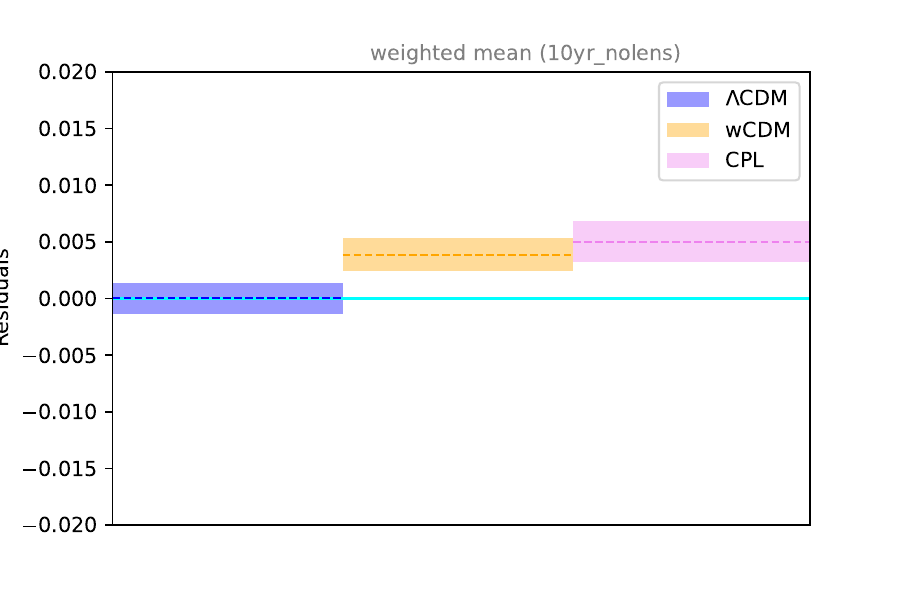}
		\caption{The $Omh^2(z_i,z_j)$ two point diagnostic residuals $R_{Omh^{2}}(z_i,z_j)$ displayed as the weighted mean for $\Lambda$CDM, $\omega$CDM and CPL model at the observed level of GW standard siren DECIGO with dashed lines and different color bands denoting $1\sigma$ confidence level. From upper to lower and left to right means 1-year and 10-year results with and without lensing.}
		\label{Fig4}
	\end{center}
\end{figure*}

\begin{figure*}
	\centering
	\includegraphics[width=0.49\linewidth]{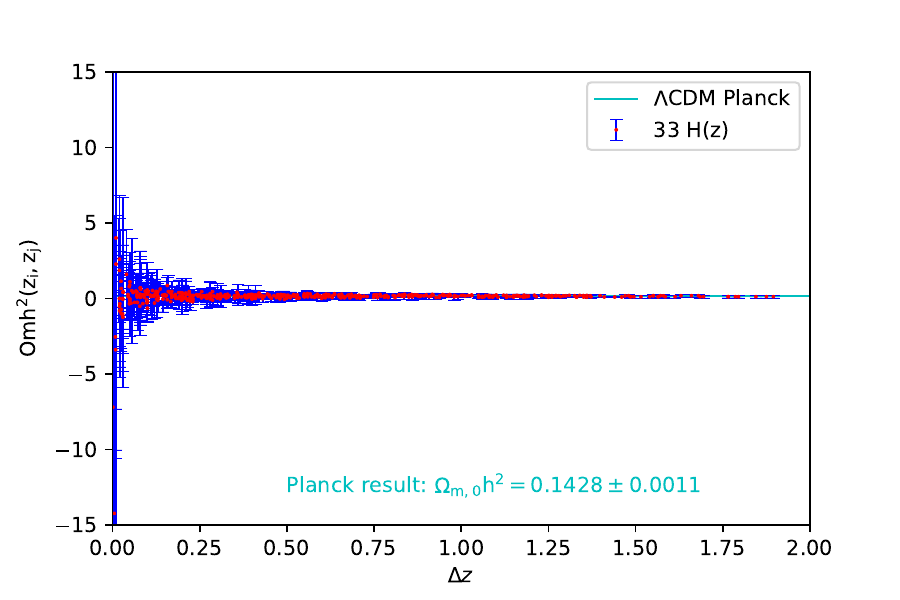}
	\includegraphics[width=0.49\linewidth]{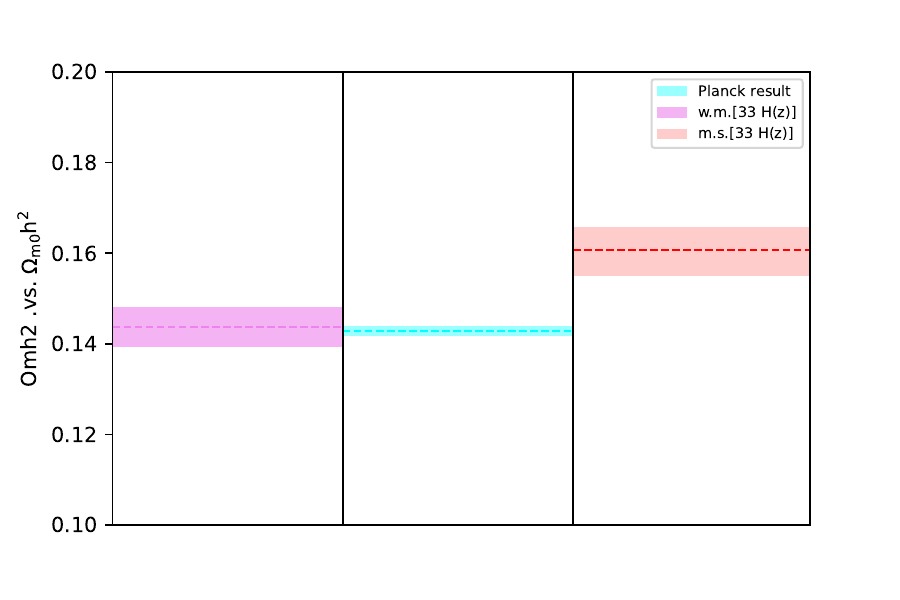}
	\caption{The $Omh^2(z_i, z_j)$ two point diagnostic calculated on the observational 33 CC $H(z)$ data. The left panel displays all 528 pairs, where the red dots with blue bars represent $Omh^2(z_i, z_j)$ and their 1-$\sigma$ confidence level. The right panel presents the statistical findings using weighted mean and median statistics.  The cyan line and bands indicate the values for which the two-point diagnostic is expected to be equal to within the $\Lambda$CDM model: $\Omega_{m,0}h^{2}=0.1428\pm0.0011$. }
	\label{Fig5}
\end{figure*}

\begin{figure*}
	\centering
	\includegraphics[width=0.49\linewidth]{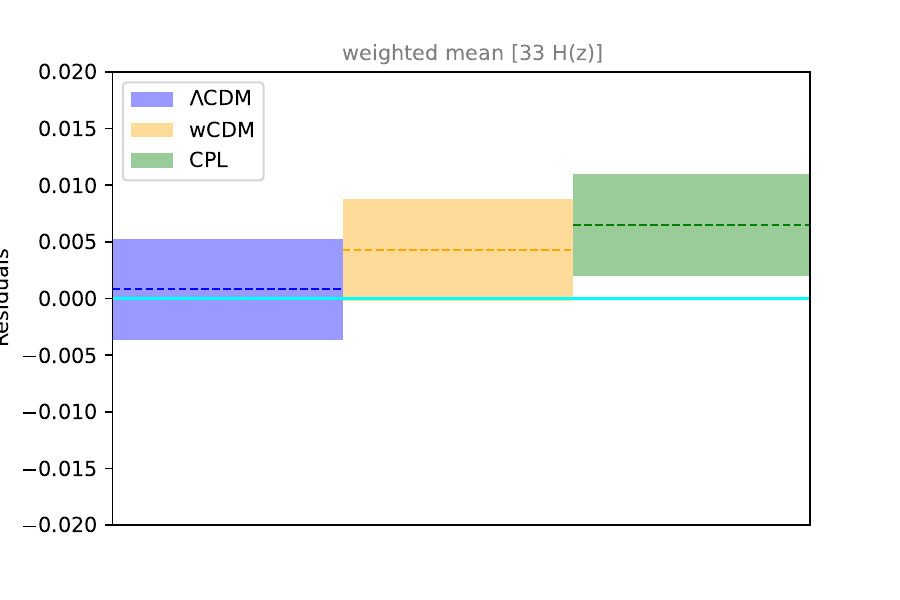}
	\includegraphics[width=0.49\linewidth]{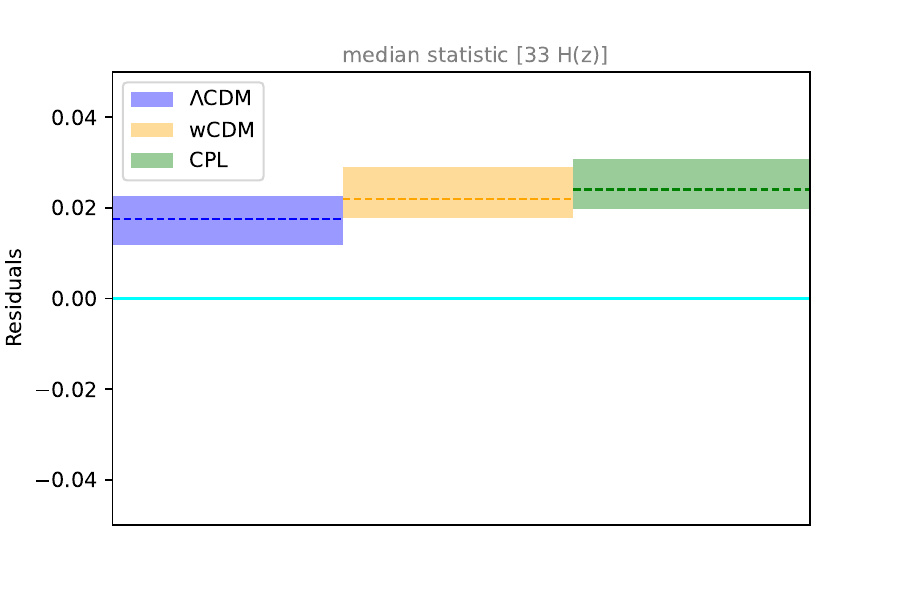}
	\caption{The $Omh^2(z_i,z_j)$ two point diagnostic residuals $R_{Omh^{2}}(z_i,z_j)$ displayed as the weighted mean (left panel) and median statistics (right panel) for $\Lambda$CDM, $\omega$CDM and CPL model at the observed level of cosmic chronometer $H(z)$ with dashed lines and different color bands denoting $1\sigma$ confidence level. }
	\label{Fig6}
\end{figure*}

\begin{table*}
	\centering
	\caption{The Weighted Mean (W.M.) and Median Statistics (m.s.) of $Omh^2(z_i,z_j)$ Two-point Diagnostics and  Residuals calculated for observational 33 $H(z)$ CC data.
	}
	\begin{tabular}{c c c c c c c c} \hline\hline
		$Omh^2(z_i,z_j)_{(w.m.)}$   &  $|N_{\sigma}|<1$  
		&  $R_{(w.m.)(\Lambda CDM)}$   &  $|N_{\sigma}|<1$  
		&  $R_{(w.m.)(\omega CDM)}$    &  $|N_{\sigma}|<1$  
		&  $R_{(w.m.)(CPL)}$           &  $|N_{\sigma}|<1$  
		\\ \hline
		$0.144\pm0.0045$  & $84.09\%$
		&  $0.0008\pm0.0045$ & $84.28\%$
		&  $0.0043\pm0.0045$ & $84.47\%$
		&  $0.0065\pm0.0046$ & $84.66\%$
		\\ \hline\hline
		$Omh^2(z_i,z_j)_{(m.s.)}$   &  $|N_{\sigma}|<1$  
		&  $R_{(m.s.)(\Lambda CDM)}$   &  $|N_{\sigma}|<1$  
		&  $R_{(m.s.)(\omega CDM)}$    &  $|N_{\sigma}|<1$  
		&  $R_{(m.s.)(CPL)}$           &  $|N_{\sigma}|<1$  
		\\ \hline
		$0.161^{+0.0051}_{-0.0057}$  & $88.64\%$
		&  $0.018^{+0.0051}_{-0.0057}$  & $88.64\%$
		&  $0.022^{+0.0071}_{-0.0042}$  & $89.02\%$
		&  $0.024^{+0.0067}_{-0.0044}$  & $89.02\%$
		\\ \hline\hline
	\end{tabular}
	\label{Table2}
\end{table*}

The inverse variance weighting as the most straightforward and popular way of summarizing multiple measurements is adopted to give the weighted mean of the individual values considered. The weighted mean formula for $Omh^2(z_i, z_j)$ two point diagnostic can be written as
\begin{equation}\label{eq:Omh2wm}
	Omh^2_{(w.m.)}=\frac{\sum^{n-1}_{i=1}\sum^{n}_{j=i+1}Omh^2(z_i,z_j)/\sigma^2_{Omh^2,ij}}{\sum^{n-1}_{i=1}\sum^{n}_{j=i+1}1/\sigma^2_{Omh^2,ij}},
\end{equation}
and its variance is given by
\begin{equation}\label{eq:Omh2wmsig}
	\sigma^2_{Omh^2_{(w.m.)}}= \left( \sum^{n-1}_{i=1}\sum^{n}_{j=i+1}1/\sigma^2_{Omh^2,ij} \right)^{-1},
\end{equation}
where $\sigma^2_{Omh^2,ij}$ can be expressed as
\begin{align}\label{eq:Omh2sig}
	\sigma^2_{Omh^2,ij}=\frac{4 \left( h^2(z_i)\sigma^2_{h(z_i)}+h^2(z_j)\sigma^2_{h(z_j)} \right)}{\left( (1+z_i)^3-(1+z_j)^3 \right)^2}.
\end{align}
In Fig.~\ref{Fig3}, one can see the weighted mean of $Omh^2(z_i,z_j)$ calculated from simulated $H(z)$ data. The blue and green dashed line means the weighted mean of $Omh^2(z_i,z_j)$ surrounded by colour bands denoting 68\% confidence regions. The darker and wider bands from the left to the right means the simulated $H(z)$ data for 1-year and 10-year observations with lensing effects, respectively. The lighter and narrower bands located at each center display the results without considering the lensing effects. As a comparison, the \textit{Planck} results of $\Omega_{m,0}h^2_{(Planck)}=0.1428\pm0.0011$ is shown by the cyan line and bands \citep{Planck2018}. Our results demonstrates the well consistency between the simulation and our prediction. We summarize the results in Table~\ref{Table1}. Although the weighted mean values are similar in all cases (with/without lensing and 1-year and 10-year observations). But what we should be more concerned about is their uncertainty, with the magnitude of $10^{-3}$. Specifically, if the observation time is the same, the uncertainty with regard to lensing is about three times greater than that without lensing. Focusing on the cases without considering the lensing effect, $\Delta_{W.M.10yr+withoutlens}=0.00114$, which means
the uncertainty of 10 year observations decreases to 32$\%$ of the 1 year observation. Of particular concern is that the uncertainty can be compared to the Planck2018 result ($\sim0.0011$) if observation lasts for 10 years (ignoring the lensing effect).

Another issue that should be discussed is the comparison between our results based on the GW standard sirens from DECIGO and other models of dark energy. More specifically, we focus on the $\omega$CDM and CPL parameterizations of dark energy. For those parameterizations, the theoretical values of $Omh^2(z_i,z_j)$ should be calculated with corresponding model parameters. Here, we use the parameters of $\Omega_{m,0}=0.299\pm0.007$ and $\omega=-1.047\pm0.038$ for $\omega$CDM and $w_0=-1.007\pm0.089$ and $w_a=-0.222\pm0.407$ for CPL parameterization from the SNe Ia and CMB constraints combined with BAO and local $H_0$ measurements \citep{Scolnic2018}. We calculated the residuals $R_{Omh^{2}}(z_i,z_j)=Omh^{2}(z_i,z_j)-Omh^{2}(z_i,z_j)_{th}$ and summarized the residuals as the weighted mean in Fig.~\ref{Fig4} and Table~\ref{Table1}. In each panel, the left blue one denotes the $\Lambda$CDM model, the middle yellow one denotes the $\omega$CDM model and the right violet one denotes the CPL parameterization. The left two panels are related to the cases when lensing effect has been considered and the right two panels are related to the case of without considering lensing. As can be seen from Fig.~\ref{Fig4}, the residuals of $\Lambda$CDM model are more closer to zero than that of $\omega$CDM and CPL model, which means that the $\Lambda$CDM model is more supported by $H(z)$ from the GW standard siren DECIGO under the $Omh^2(z_i,z_j)$ two-point diagnostics. In other words, if we do not consider lensing effects (meaning that the lensing effect was corrected), we can differentiate the $\Lambda$CDM model from others clearly even for 1 year observations. In the case when lensing uncertainty is taken into account, these three models can be differentiated only after more than 3 years of DECIGO observations.

Finally, one could wonder the performance of other models of dark energy in the EM domain with the latest $H(z)$ measurements. \citet{Jimenez2002} proposed a cosmological model-independent technique to directly measure $H(z)$, which is known as differential age (DA) or cosmic chronometer (CC) method. In this method, the Hubble parameter $H(z)$ can be obtained as:
\begin{equation}
	H(z)=-\frac{1}{1+z}\frac{dz}{dt},
\end{equation}
where $z$ is the cosmological redshift and $dz/dt$ can be directly obtained by measuring the age difference $\Delta t$ between two galaxies, which are massive and evolving passively on a timescale larger than their age difference and differ in redshifts by $\Delta z$. Some features of their spectrum such as the $D4000$ break enable us to measure the age difference of such galaxies.
Another method to obtain the $H(z)$ is the radial baryon acoustic oscillations (BAO) method \citep{Blake2012,Delubac2015,Blomqvist2019} based on the value of the sound horizon at the drag epoch $r_d$. It should be noted that this approach assumes a priori the cosmological model, which makes such measurements model-dependent \citep{Li2016}. In this paper, we utilize the newest 33 CC $H(z)$ measurements with redshift up to $\sim2$, wherein 32 measurements have been extensively employed in previous studies \citep{Borghi2022,Wu2023} while 1 measurement has been updated \citep{Tomasetti2023}.

The results of 528 pairs of $Omh^2(z_i,z_j)$ are shown in Fig.~\ref{Fig5}, where red points denote the central values and blue bars are corresponding 68$\%$ uncertainties. It is clearly visible that almost all of the reconstructed $Omh^2(z_i,z_j)$ are consistent with the theoretical values from $\Lambda$CDM in the $1\sigma$ confidence level. In order to improve the robustness of the results, we also used the weighted mean statistics. Table~\ref{Table2} lists the results of this statistical method. For all $Omh^2(z_i,z_j)$ pairs, the result of weighted mean $0.1438\pm0.0045$ is consistent with Planck2018 result $0.1428\pm0.0011$. Compared to our previous results using simulated $H(z)$ data in the GW domain, we find that if we ignore the lensing effect, the accuracy of one year observation is comparable to the current Hubble parameter observations. Even taking into account the lensing effect, the current accuracy can be achieved after more than 3-year observations. Moreover, the assumption underlying the utilization of the weighted mean approach is that the error distribution must conform to a Gaussian distribution, we adopt the methodology used by \citet{Chen2003} to test the Gaussianity of error distributions. In this regard, it is necessary to compute the number of standard deviations $N_{\sigma}$ by which a measurement deviates from its central value. The result for the weighted mean is 84.09\%, clearly indicating that the error does not follow a Gaussian distribution. Additionally, it should be noted that the percentage of measurements with $|N_\sigma|<1$ should ideally be equal to 68.3\% for Gaussian Distribution. Therefore, we further use the median statistic approach for analysis. When making a total number of $N$ measurements, one might naturally expect that there is a 50\% chance that each measurement is higher/lower than the true median. Therefore, the probability that $n$-th observation is higher than the median follows the binomial distribution: $P=2^{-N}N!/[n!(N-n)!]$ \citep{Gott2001}. Similarly, we can define the 68.3$\%$ confidence interval with median statistics. The obtained result of the median statistics $0.161^{+0.0051}_{-0.0057}$ is incompatible with Planck2018 result derived from the $\Lambda$CDM model. The results derived from aformentioned two statistical methodologies are illustrated in Fig.~\ref{Fig5} and Table~\ref{Table2}. We present weighted mean of residuals in Fig.~\ref{Fig6} and the corresponding values are shown in Table.~\ref{Table2}. In principle, the residual should be zero. For comparison between different dark energy parameterizations, the standard $\Lambda$CDM model still performed the best which is consistent with our previous results. As for the $\omega$CDM or CPL models, the residuals summarized in the median statistics scheme present some deviation from the expected value of zero. One can make a conclusion from Fig.~\ref{Fig6} that the residuals $Omh^2(z_i,z_j)$  for the $\Lambda$CDM model are closer to zero than that of the $\omega$CDM or CPL models in such statistical method we use \citep{Zheng2016}.

\section{Conclusion}\label{sec:con}

The $Omh^2(z_i,z_j)$ two point diagnostics as an independent and popular screening test of the spatially flat standard $\Lambda$CDM model was used in our research. One of its advantages is that it is only related to the Hubble parameter $H(z)$ \citep{Sahni2014}.
Therefore, the sample size and the redshift range of Hubble parameters $H(z)$ are very important. Fortunately, we are in a new era of gravitational wave multi-messenger astronomy, which gives us more opportunities and possibilities to test various physical theories, not just $\Lambda$CDM model. DECIGO as the future space-based GW detector is very promising since it would be able to detect a large number of sources (neutron star binaries) distributed deeply enough at higher redshift. More importantly, the neutron star binaries would be clean GW sources, which can provide the  luminosity distance with less systematics \citep{Nishizawa2011}.
In consequence, we would have a chance to obtain the Hubble parameters $H(z)$ directly by using the dipole of the  luminosity distance from the DECIGO. Our results don't show a significant deviation compared with the \textit{Planck} results, which means that the $\Lambda$CDM model (assumed in the simulations) is supported by the $Omh^2(z_i,z_j)$  two point diagnostics with $H(z)$ expansion rates from the second-generation space-based GW detector, DECi-hertz Interferometer Gravitational-wave Observatory (DECIGO). In particular, in the framework of DECIGO,  the $Omh^2(z_i,z_j)$  two point diagnostics are expected to be constrained with the precision of $10^{-3}$. And the uncertainty (0.00114) can be comparable with the Planck2018 result (0.0011) if observed for 10 years (ignoring the lensing effect). In addition, we have also discussed the performance of $\omega$CDM and CPL models. Residual results show that there is a certain deviation seen in the data from 10 years of observation. In other words, we can differentiate these models with standard $\Lambda$CDM cosmological model from 10-yr observation cycle when the uncertainty from lensing is considered or from 3-year observation data when the lensing is mitigated.

For comparison, the newest measurements of 33 cosmic chronometers with redshift up to $\sim2$ were also considered, which raise the possibility of testing cosmological models in the EM domain. We found no significant inconsistencies within the $68\%$ confidence level in the scheme of the $Omh^2(z_i,z_j)$ two point diagnosis and weighted mean statistical method, which is consistent with Planck2018 result and our previous results in GW domain. If we ignore the lensing effect, the accuracy of 1-year observation is comparable to the current Hubble parameter observations. Even taking into account the lensing effect, the current accuracy can be achieved after more than 3-year observations. For different dark energy models, it is important to note that the residuals of the standard cosmological model are closer to zero than its immediate extensions, which turned to be that $\Lambda$CDM model still performs better under the $Omh^2(z_i,z_j)$ two-point diagnosis with the CC $H(z)$ measurements.

In conclusion, the precision of Hubble parameters derived from gravitational waves possess significant potential in discriminating between $\Lambda$CDM and parametric models. We can differentiate these models with $\Lambda$CDM from 10-yr observation cycle when the uncertainty from lensing is considered or from 3-year observation data when the lensing is mitigated. To discuss the accuracy of Hubble parameters for distinguishing cosmological models, the most recent Hubble parameter comes from cosmic chronometer with $z\sim2$ is employed to conduct calculations based on the two-point diagnosis of $Omh^2(z_i,z_j)$ and the results demonstrate consistency with the outcomes from Planck's $\Lambda$CDM model, aligning with our prior research findings. We would like to stress that this test is independent and does not rely on any cosmological assumptions. Most importantly, it is a new attempt to test the theory using the Hubble parameter measured directly by gravitational waves.

\section*{Acknowledgments}
We would like to thank Jingzhao Qi and Masamune Oguri for his helpful discussions. This work was supported by the National Natural Science Foundation of China under Grant Nos. 12021003, 12203009, 11690023, 11633001 and 11920101003; the Strategic Priority Research Program of the Chinese Academy of Sciences, Grant No. XDB23000000. Y.T. Liu was supported by the Interdiscipline Research Funds of Beijing Normal University under Grant No. BNUXKJC2017 and China Scholarship Council under Grant No. 202106040084. X.G. Zheng was supported by the National Natural Science Foundation of China under Grant No .12103036 and the research funds of Wuhan Polytechnic University.

\section*{Data Availability Statement} This manuscript has associated data in a data repository. [Authors'Comment: The data underlying this paper will be shared on reasonable request to the corresponding author.]

\end{document}